\documentclass[aps,nofootinbib,preprint]{revtex4}
\usepackage{color}

\begin{document}

%\preprint{hep-th/yymmnnn}

\title{On Holographic Dual of the Dyonic Reissner-Nordstr\"om \\ Black Hole}

\author{Chiang-Mei Chen} \email{cmchen@phy.ncu.edu.tw}
\affiliation{Department of Physics and Center for Mathematics and Theoretical Physics,
National Central University, Chungli 320, Taiwan}

\author{Ying-Ming Huang} \email{y.m.huang26@gmail.com}
\affiliation{Department of Physics, National Central University, Chungli 320, Taiwan}

\author{Jia-Rui Sun} \email{jrsun@phy.ncu.edu.tw}
\affiliation{Department of Physics, National Central University, Chungli 320, Taiwan}

\author{Ming-Fan Wu} \email{93222036@cc.ncu.edu.tw}
\affiliation{Department of Physics, National Central University, Chungli 320, Taiwan}

\author{Shou-Jyun Zou} \email{sgzou2000@gmail.com}
\affiliation{Department of Physics, National Central University, Chungli 320, Taiwan}

\date{\today}

%% REVTEX4
%\maketitle

\begin{abstract}
It is shown that the hidden conformal symmetry, namely $SO(2,2) \sim
SL(2,R)_L \times SL(2,R)_R$ symmetry, of the non-extremal
dyonic Reissner-Nordstr\"om black hole can be probed by a charged
massless scalar field at low frequencies. The existence of such
hidden conformal symmetry suggests that the field theory
holographically dual to the 4D Reissner-Nordstr\"om black hole
indeed should be a 2D CFT. Although the associated AdS$_3$ structure
does not explicitly appear in the near horizon geometry, the primary
parameters of the dual CFT$_2$ can be exactly obtained without the
necessity of embedding the 4D Reissner-Nordstr\"om black hole into
5D spacetime. The duality is further supported by comparing the
absorption cross sections and real-time correlators obtained from
both the CFT and the gravity sides.
\end{abstract}

%% REVTEX4
%\pacs{}

\maketitle
\tableofcontents

%%%%%%%%%%%%%%%%%%%%%%%%%%%%%%%%%%%%%%%%%%%%%%%%%%%%%%%%%%%%%%%%%%%%%%
\section{Introduction}
%%%%%%%%%%%%%%%%%%%%%%%%%%%%%%%%%%%%%%%%%%%%%%%%%%%%%%%%%%%%%%%%%%%%%%
Searching for the quantum gravity descriptions of black holes has a
long story. The conceivable first explicit example is the
calculation of the central charge of 2D CFT by analyzing the
asymptotic symmetries of the asymptotically AdS$_3$ spacetimes,
namely the AdS$_3$/CFT$_2$ correspondence~\cite{Brown:1986nw}.
Later, based on the ideas of the holographic
principle~\cite{'tHooft:1993gx, Susskind:1994vu} and its first
realization in string theory, i.e. AdS$_5$/CFT$_4$
correspondence~\cite{Maldacena:1997re, Gubser:1998bc,
Witten:1998qj}, much success have been made along this direction
especially for the recent progresses initiated by studying the
quantum gravity descriptions of the Kerr black hole, i.e. the
Kerr/CFT correspondence~\cite{Guica:2008mu}, together with its
various applications and extensions~\cite{Dias:2009ex,
Matsuo:2009sj, Bredberg:2009pv, Amsel:2009pu, Hartman:2009nz,
Castro:2009jf, Cvetic:2009jn, Hartman:2008pb, Garousi:2009zx,
Chen:2009ht, Chen:2010bs, Hotta:2008xt, Lu:2008jk, Azeyanagi:2008kb,
Chow:2008dp, Azeyanagi:2008dk, Nakayama:2008kg, Isono:2008kx,
Peng:2009ty, Chen:2009xja, Loran:2009cr, Ghezelbash:2009gf,
Lu:2009gj, Amsel:2009ev, Compere:2009dp, Krishnan:2009tj,
Hotta:2009bm, Astefanesei:2009sh, Wen:2009qc, Azeyanagi:2009wf,
Wu:2009di, Peng:2009wx, Chen:2009cg, Chen:2010ni, Becker:2010jj,
Balasubramanian:2009bg, Castro:2010vi}. In almost all the above
cases, in order to apply the technique of the AdS/CFT duality, the
background spacetime (in the extremal or near extremal limit) needs
to contain some asymptotical or/and near horizon AdS structures. The
Kerr/CFT correspondence is of no exception in the beginning, in
which the near horizon AdS$_2 \times S^1$ geometry (with $SL(2, R)_R
\times U(1)_L$ symmetry) of the (near) extremal Kerr black hole
plays an essential role in obtaining the central charge of the dual
2D CFT.

The $S^1$ or $U(1)$ bundle of the Kerr black hole comes from the
rotation, thus for non-rotating black holes such as the
Reissner-Nordstr\"om (RN) black hole which only has the near horizon
(near) extremal AdS$_2$ structure, the Kerr/CFT approach does not
work directly. One possible way is to uplift the 4D RN black hole
into 5D spacetime and let the extra dimension be a compactified
$S^1$ circle (with radius $\ell$), then the left hand central charge
$c_L = 6 Q^3/\ell$ of the dual CFT can be
calculated~\cite{Hartman:2008pb, Garousi:2009zx, Chen:2010bs}.
Another way is to reduce the 4D RN black hole into two dimensions
and study the dynamics of the stress tensor and current of the 2D
effective theory, which results, with a suitable choice of an
undetermined normalization factor, the right hand central charge
$c_R = 6 Q^2$ of the dual CFT~\cite{Chen:2009ht}. The first picture
shows that the near horizon (near) extremal uplifted 5D RN black
hole has the (warped) AdS$_3$/CFT$_2$ description. While the second
picture indicates that the near horizon (near) extremal 4D RN black
hole has the AdS$_2$/CFT$_1$ description. Hence it is suggested
in~\cite{Chen:2009ht, Chen:2010bs} that the 4D and 5D RN black holes
may serve as a possible example to study the relationship between
the AdS$_3$/CFT$_2$ and AdS$_2$/CFT$_1$ dualities. On the other
hand, the Kerr/CFT correspondence is further conjectured to work
even in the generic non-extremal case based on the fact that the
hidden conformal symmetry of the Kerr black hole can be probed by
the external fields in the low frequency limit~\cite{Castro:2010fd}.
In this method, the $U(1)$ bundle also plays a vital important part
in constructing the Casimir operator of the $SL(2,R)_L \times
SL(2,R)_R$ Lie algebra. We soon show that the hidden conformal
symmetry can be found also for the 5D RN black hole, which indicates
that the generic non-extremal 5D RN black hole is dual to a 2D
CFT~\cite{Chen:2010as}. For other related works with the hidden
conformal symmetry, see~\cite{Krishnan:2010pv, Wang:2010qv,
Rasmussen:2010sa, Chen:2010xu, Li:2010ch, Chen:2010zw,
Krishnan:2010df, Rasmussen:2010xd, Becker:2010dm, Chen:2010bh,
Wang:2010ic}.

However, whether the 4D RN black hole is dual the 2D or 1D CFTs
remains an interesting problem. In the present paper we find that,
even for the 4D non-extremal RN black hole, there should also exist
a dual CFT$_2$ description, without the necessity of uplifting it
into higher dimensions. What is more, the dyonic RN solution does
not have a trivial embedding into five dimensional spacetime, so it
is more natural to study the CFT dual directly in 4 dimensions.
Although the techniques we used is similar to those used in the Kerr
black hole case, the key point is that the $U(1)$ bundle of the 4D
RN black hole can be probed by a charged scalar field but not just a
neutral one. This is not surprising since the $U(1)$ symmetry is
actually from the gauge symmetry of the background electromagnetic
field. In practice, the gauge potential provides, via the coupling
with a charged scalar field, the $U(1)$ fibration over the AdS$_2$
base manifold to form an AdS$_3$ structure, as we expect. Then the
generic 4D dyonic RN black hole is conjectured to be dual to a 2D
CFT with $T_L = \frac{(r_+^2 + r_-^2) \ell}{4 \pi Q (Q^2 + P^2)}$
and $T_R = \frac{(r_+^2 - r_-^2) \ell}{4 \pi Q (Q^2 + P^2)}$, and
$c_L = c_R = \frac{6 Q (Q^2 + P^2)}{\ell}$. Besides the matching of
the microscopic and macroscopic entropies
$$
S_{CFT} = \frac{\pi^2}{3} (c_L T_L + c_R T_R) = \pi r_+^2 = S_{BH},
$$
we also find that, the absorption cross section of the scalar field
calculated from the gravity side matches with those of its dual
operators in the 2D CFT exactly, and the real time correlators
obtained from the gravity and the CFT sides are in good agreement up
to some normalization factors depending on the charge of the probe
field, which give further supports to the AdS$_3$/CFT$_2$ picture of
the 4D RN black hole. Based on this work, we would like to
point out that a dimensional reduction in the gravity side, namely
from uplifting 4D RN black hole to its 5D counterpart, can not
provide connection between the dual CFT$_2$ and CFT$_1$. Actually,
no matter for 4D RN or 5D RN black holes, their corresponding CFTs
should be two dimensional.

The outline of this paper is as follows. We first review the basic
properties of the dyonic RN black hole and then analyze the
scattering process of a charged scalar field propagating in this
background in Section II. In Section III, we illustrate how the
hidden conformal symmetry of the 4D RN black hole is probed by a
charged massless scalar field in the low frequency limit, and
consequently, how the dual CFT$_2$ description appears. In Sections
IV and V, we further verify the AdS$_3$/CFT$_2$ picture by comparing
the absorption cross sections and the real time correlators
calculated from both the gravity and the CFT$_2$ sides. Then we give
the conclusion in Section VI. In Appendix A, we list the symmetry
and the Casimir operator in the AdS$_3$ spacetime.

%%%%%%%%%%%%%%%%%%%%%%%%%%%%%%%%%%%%%%%%%%%%%%%%%%%%%%%%%%%%%%%%%%%%%%
\section{Charged Scalar Field in the Dyonic RN Black Hole}
%%%%%%%%%%%%%%%%%%%%%%%%%%%%%%%%%%%%%%%%%%%%%%%%%%%%%%%%%%%%%%%%%%%%%%
We begin by studying a charged probe scalar field propagating in the
dyonic RN black hole background. The dyonic RN black holes are the
spherically symmetric charged solutions of the four-dimensional
Einstein-Maxwell theory
\begin{equation}
I_4 = \frac1{16 \pi} \int d^4x \sqrt{-g} \left( R - F_{[2]}^2 \right),
\end{equation}
it is characterized by three parameters: one is the mass $M$ and the
other two are electric charge $Q$ and magnetic charge $P$ respectively,
\begin{eqnarray}\label{RN}
ds^2 &=& - f(r) dt^2 + \frac{dr^2}{f(r)} + r^2 d\Omega_2^2,
\nonumber\\
A_{[1]} &=& - \frac{Q}{r} dt + P \cos\theta d\phi,
\end{eqnarray}
where
\begin{equation}
\qquad f(r) = 1 - \frac{2 M}{r} + \frac{Q^2 + P^2}{r^2}.
\end{equation}
The corresponding outer/inner horizon radius $r_\pm$, chemical
potential $\Phi_H$, Hawking temperature $T_H$ and black hole entropy
$S_{BH}$ are
\begin{eqnarray}
r_\pm &=& M \pm \sqrt{M^2 - Q^2 - P^2},
\nonumber\\
\Phi_H &=& \frac{Q}{r_+},
\nonumber\\
T_H &=& \frac1{4 \pi} \frac{r_+ - r_-}{r_+^2},
\nonumber\\
S_{BH} &=& \frac{A_+}4 = \pi r_+^2.
\end{eqnarray}

For a probe charged massive scalar field $\Phi$ of mass $\mu$ and
electric charge $q$, which is minimally coupled to the dyonic RN
black hole background, its corresponding Klein-Gordon (KG) equation
\begin{equation} \label{KG}
(\nabla_\alpha - i q A_\alpha) (\nabla^\alpha - i q A^\alpha) \Phi +
\mu^2 \Phi = 0,
\end{equation}
can be simplified by assuming the following mode expansion of the
scalar field
\begin{equation} \label{APhi}
\Phi(t, r, \theta, \phi) = \mathrm{e}^{-i \omega t + i m \phi} S(\theta) R(r),
\end{equation}
and, accordingly the KG equation reduces to two decoupled radial and
angular equations by separation of variables:
\begin{eqnarray}\label{radial}
\partial_r (\Delta \partial_r R) + \left[ \frac{(\omega r - q Q)^2 r^2}{\Delta} - \mu^2 r^2 - \lambda_l \right] R &=& 0,
\\
\frac1{\sin\theta} \partial_\theta (\sin\theta \partial_\theta S_l) + \left[ \lambda_l - \frac{(m - q P)^2}{\sin^2\theta} \right] S_l &=& 0,
\end{eqnarray}
where $\Delta = r^2 f = (r - r_+) (r - r_-)$. For the massless
scalar field, $\mu = 0$, the radial equation~(\ref{radial}) can be
further reformulated in the following symmetric form
\begin{eqnarray}\label{radial2}
&& \partial_r (\Delta \partial_r R) + \Biggl[ \frac{\left( r_+^2
\omega - q Q r_+ \right)^2}{(r - r_+)(r_+ - r_-)} - \frac{\left(
r_-^2 \omega - q Q r_- \right)^2}{(r - r_-)(r_+ - r_-)}
\nonumber\\
&& \qquad + (\omega r + 2 \omega M - q Q)^2 - \omega^2 (2 M r + Q^2 + P^2) \Biggr] R = \lambda_l R.
\end{eqnarray}
In order to reveal the 2D conformal symmetry in the radial
part equation of motion~(\ref{radial2}), one should impose proper
conditions on the probe scalar field~(\ref{APhi}) such that the
potential terms in the second line of Eq.(\ref{radial2}) can be
neglected. These conditions are: (i) low frequency: $\omega M \ll 1$
(consequently $\omega Q \ll 1$ and $\omega P \ll 1$), (ii) small
electric charge: $q Q \ll 1$ and $q P \ll 1$, (iii) near region:
$\omega r \ll 1$. Note that the condition $q P \ll 1$ simplifies the
angular equation in the way that the separation constant should take
the standard value $\lambda_l = l (l + 1)$, and the solutions for
$S_l$ are just the standard spherical harmonic functions.

%%%%%%%%%%%%%%%%%%%%%%%%%%%%%%%%%%%%%%%%%%%%%%%%%%%%%%%%%%%%%%%%%%%%%%
\section{Hidden Conformal Symmetry}
%%%%%%%%%%%%%%%%%%%%%%%%%%%%%%%%%%%%%%%%%%%%%%%%%%%%%%%%%%%%%%%%%%%%%%
The reduced radial equation~(\ref{radial2}), after imposing the
suitable conditions at the end of the Section II, is
\begin{equation}\label{scalarEQ}
\partial_r (\Delta \partial_r R) + \left[ \frac{r_+^4 (\omega + A_+ q)^2}{(r - r_+)(r_+ - r_-)} - \frac{r_-^4 (\omega + A_- q)^2}{(r - r_-)(r_+ - r_-)} \right] R = l (l + 1) R,
\end{equation}
where $A_\pm = - Q / r_\pm$ are the gauge potentials at the
outer/inner horizons. Defining the covariant derivative operators as
$D_\pm = \partial_t - i q A_\pm$, then we can have the relations $D_\pm \Phi = - i (\omega
+ q A_\pm) \Phi$. Consequently, we find that the reduced radial
equation~(\ref{scalarEQ}) matches with the Casimir operator of the
$SL(2,R)_L \times SL(2,R)_R$ Lie algebra~(\ref{Casimir}) by imposing
the following relations
\begin{eqnarray} \label{OPrelation}
- r_+^2 \, D_+ &=& (r_+ - r_-) \left(\frac{T_L + T_R}{4 \mathcal{A}} \partial_t - \frac{n_L + n_R}{4 \pi \mathcal{A}} \partial_\chi\right),
\nonumber\\
- r_-^2 \, D_- &=& (r_+ - r_-) \left(\frac{T_L - T_R}{4 \mathcal{A}} \partial_t - \frac{n_L - n_R}{4 \pi \mathcal{A}} \partial_\chi\right),
\end{eqnarray}
where $\mathcal{A}$ is defined in~(\ref{deltaA}). Here a suitable
sign is chosen to ensure the temperatures to be positive. In the
above identifications, we have introduced an operator
$\partial_\chi$, which can be naturally considered to act on the
``internal'' $U(1)$ symmetry space of the probe electrically charged
scalar field and its eigenvalue gives the electric charge of the
probe field such as $\partial_\chi \Phi = i \ell q \Phi$, where the
parameter $\ell$ is,
in general, an arbitrary normalization factor %\footnote{This parameter
%has a natural geometrical interpretation as the radius of extra circle
%when the RN solution is considered be embedded into five-dimensional
%spacetime~\cite{Chen:2010bs}.}
and it should not affect entropy of the dual 2D CFT. Thus, the covariant
derivatives can be rewritten as $D_\pm = \partial_t - (A_\pm/\ell)
\partial_\chi = \partial_t + (Q/\ell r_\pm) \partial_\chi$ and the
relations~(\ref{OPrelation}) decompose into four algebraic equations
\begin{eqnarray}
- r_+^2 = \frac{(r_+ - r_-) (T_L + T_R)}{4 \mathcal{A}}, &\quad& - r_+^2 \, A_+ = \ell \frac{(r_+ - r_-) (n_L + n_R)}{4 \pi \mathcal{A}},
\nonumber\\
- r_-^2 = \frac{(r_+ - r_-) (T_L - T_R)}{4 \mathcal{A}}, &\quad& - r_-^2 \, A_- = \ell \frac{(r_+ - r_-) (n_L - n_R)}{4 \pi \mathcal{A}}.
\end{eqnarray}
The temperatures of the left hand- and right hand- CFTs, i.e. $T_L$
and $T_R$, then can be straightforwardly computed
\begin{eqnarray}
T_L &=& \frac{(r_+^2 + r_-^2) \ell}{4 \pi Q r_+ r_-} = \frac{(r_+ + r_-) M \ell - (Q^2 + P^2) \ell}{2 \pi Q (Q^2 + P^2)},
\\
T_R &=& \frac{(r_+^2 - r_-^2) \ell}{4 \pi Q r_+ r_-} = \frac{(r_+ - r_-) M \ell}{2 \pi Q (Q^2 + P^2)},
\end{eqnarray}
as well as the two other quantities $n_L$ and $n_R$
\begin{equation}
n_L = - \frac{r_+ + r_-}{4 r_+ r_-} = - \frac{r_+ + r_-}{4 (Q^2 + P^2)}, \qquad n_R = - \frac{r_+ - r_-}{4 r_+ r_-} = - \frac{r_+ - r_-}{4 (Q^2 + P^2)}.
\end{equation}

In addition, the central charges of the dual CFT corresponding to
the dyonic RN black hole have been studied both in the 5D uplifted
picture and the 2D reduced picture via analyzing the asymptotical
symmetry of the near horizon (near) extremal
geometry~\cite{Hartman:2008pb, Garousi:2009zx, Chen:2010bs,
Chen:2009ht}. As has been discussed in~\cite{Chen:2010as} that in
both of the two approaches, there is an undetermined free parameter
$\ell'$
--- may not be the same with the parameter $\ell$ at the moment ---
appearing in the temperatures and central charges of the dual CFTs.
The generic expression for the central charges are
\begin{equation}\label{cc}
c_L = c_R = \frac{6 Q (Q^2 + P^2)}{\ell'}.
\end{equation}
%Here we use $\ell'$ to label the parameter because generally $\ell'$ is not the same as $\ell$.
%However, they can be identified in the following explanation.
Note that the value of $\ell'$ in the 5D picture labels the radius
of extra dimensional circle while $\ell$ in the 4D picture also
describes the size of an ``internal'' circle. With the relation
$\partial_\chi \Phi = i \ell q \Phi$, it is clear to see that
Eq.(\ref{scalarEQ}) is exactly the same as the radial part equation
for a neutral probe scalar field scattering in the 5D uplifted RN
black hole background at low
frequencies~\cite{Chen:2010as}\footnote{This fact actually indicates
that the 4D and 5D RN black holes should dual to the same 2D CFT.}.
Therefore, we should identify $\ell' = \ell$
%If we further assume that the central charges in Eq.(\ref{cc}) don't change in the non-extremal RN
%black hole case,
and then the CFT microscopic entropy characterized by
the Cardy formula agrees with the black hole macroscopic area
entropy
\begin{equation}
S_\mathrm{CFT} = \frac{\pi^2}3 ( c_L T_L + c_R T_R ) = \pi r_+^2 = S_\mathrm{BH}.
\end{equation}
Like the Kerr black hole and the 5D RN black hole, the hidden
conformal symmetry, i.e. the $SL(2,R)_L \times SL(2,R)_R$ symmetry
of the generic non-extremal dyonic RN black hole cannot be detected
globally in the solution space of Eq.(\ref{scalarEQ}), it will break
from $SL(2,R)_L \times SL(2,R)_R$ into $U(1)_L \times U(1)_R$ due to
the periodic identification of the internal $U(1)$ symmetry
\begin{equation}
\chi \sim \chi + 2 \pi.
\end{equation}
Accordingly, the generators of the $SL(2,R)_L \times SL(2,R)_R$ Lie
algebra listed in Appendix A transform as
\begin{equation}
w^+ \sim \mathrm{e}^{4\pi^2 T_R} \, w^+, \qquad w^- \sim
\mathrm{e}^{4\pi^2 T_L} \, w^-, \qquad y \sim \mathrm{e}^{2\pi^2(T_L
+ T_R)} \, y.
\end{equation}

%%%%%%%%%%%%%%%%%%%%%%%%%%%%%%%%%%%%%%%%%%%%%%%%%%%%%%%%%%%%%%%%%%%%%%
\section{Scattering of the Probe Scalar Field}
%%%%%%%%%%%%%%%%%%%%%%%%%%%%%%%%%%%%%%%%%%%%%%%%%%%%%%%%%%%%%%%%%%%%%%
In addition to the matching of CFT and black hole entropies, the
dyonic RN/CFT$_2$ correspondence can be further checked by studying
the scattering of the probe charged scalar field in the RN black
hole background. We will see that the absorption cross section of
the scalar field is in agreement with the two point function of its
corresponding operators in the dual 2D CFT.

From the gravity side\footnote{Note that even though the calculation
of scattering amplitudes does not relate to the geometric quantities
in the gravity side, directly, it does extract the information of
the geometric background since the probe field is coupled to the
gravity.}, the near region KG equation~(\ref{scalarEQ}) can be
easily solved by introducing a new variable $z$
\begin{equation}
z = \frac{r - r_+}{r - r_-},
\end{equation}
and the general solution includes both ingoing and outgoing modes
\begin{eqnarray}
R^\mathrm{(in)} &=& z^{- i \gamma} (1 - z)^{l + 1} \, F( a, b; c; z ),
\nonumber\\
R^\mathrm{(out)} &=& z^{i \gamma} (1 - z)^{l + 1} \, F( a^*, b^*; c^*; z ),
\end{eqnarray}
in which $F( a, b; c; z )$ is the hypergeometric function and all
the coefficients are given by
\begin{eqnarray}
\gamma &=& \frac{(\omega r_+ - q Q) r_+}{r_+ - r_-},
\nonumber\\
a &=& 1 + l - i \frac{\omega (r_+^2 + r_-^2) - q Q (r_+ + r_-)}{r_+ - r_-},
\nonumber\\
b &=& 1 + l - i \left[ \omega (r_+ + r_-) - q Q \right],
\nonumber\\
c &=& 1 - i 2 \gamma.
\end{eqnarray}
An useful relation among these coefficients is
\begin{equation}\label{cab}
c - a - b = - 2 l - 1.
\end{equation}
In the matching region $r \gg M$ and $r \ll 1/\omega$ (which is
equivalent to take the limits $z \to 1$ and $1 - z \to r^{-1}$), the
ingoing mode asymptotically behaves as
\begin{equation}\label{asymR}
R^\mathrm{(in)}(r \gg M) \sim A \, r^l + B \, r^{- l - 1},
\end{equation}
where two coefficients $A$ and $B$ are
\begin{equation}
A = \frac{\Gamma(c) \Gamma(2 l + 1)}{\Gamma(a) \Gamma(b)}, \qquad B
= \frac{\Gamma(c) \Gamma(- 2 l - 1)}{\Gamma(c - a) \Gamma(c - b)}.
\end{equation}
One more additional information can be obtained from~(\ref{asymR})
is the conformal weights of the scalar field
\begin{equation}
h_L = h_R = l + 1.
\end{equation}
Hence, the coefficients $a$ and $b$ can be expressed in terms of conformal weights (real part) and two parameters $\tilde\omega_L$ and $\tilde\omega_R$ (imaginary part)
\begin{equation}
a = h_R - i \frac{\tilde\omega_R}{2 \pi T_R}, \qquad b = h_L - i \frac{\tilde\omega_L}{2 \pi T_L},
\end{equation}
where $\tilde\omega_L, \tilde\omega_R$ are composed by three sets of
the parameters: frequencies ($\omega_L, \omega_R$), charges ($q_L,
q_R$) and chemical potentials ($\mu_L, \mu_R$)
\begin{eqnarray}
\tilde\omega_L = \omega_L - q_L \mu_L, &\quad& \omega_L = \frac{\ell \omega (r_+ + r_-)(r_+^2 + r_-^2)}{2 Q r_+ r_-}, \quad q_L = q, \quad \mu_L = \frac{\ell (r_+^2 + r_-^2)}{2 r_+ r_-},
\nonumber\\
\tilde\omega_R = \omega_R - q_R \mu_R, &\quad& \omega_R = \frac{\ell \omega (r_+ + r_-)(r_+^2 + r_-^2)}{2 Q r_+ r_-}, \quad q_R = q, \quad \mu_R = \frac{\ell (r_+ + r_-)^2}{2 r_+ r_-}.
\end{eqnarray}
The essential part of the absorption cross section of the charged
scalar field can be read out directly from the coefficient $A$ in
Eq.(\ref{asymR}), namely
\begin{equation}\label{Pabs}
P_\mathrm{abs} \sim |A|^{-2} \sim \sinh( 2\pi \gamma) \, \left| \Gamma(a) \right|^2 \, \left| \Gamma(b) \right|^2.
\end{equation}

In order to compare the absorption cross section~(\ref{Pabs}) with
the two-point function of the operator dual to the probe scalar
field, one needs to identify the conjugate charges, $\delta E_L$ and
$\delta E_R$, defined by
\begin{equation}
\delta S_{CFT} = \frac{\delta E_L}{T_L} + \frac{\delta E_R}{T_R},
\end{equation}
from the first law of black hole thermodynamics
\begin{equation}
\delta S_{BH} = \frac1{T_H} \delta M - \frac{\Phi_H}{T_H} \delta Q.
\end{equation}
Here the magnetic charge is fixed in the first law, i.e. $\delta P =
0$, since the probe scalar field carries only the electric charge,
therefore, it can not probe the the background magnetic charge and
hence cannot take the magnetic charge away from the black hole.
Finally one can get the conjugate charges via $\delta S_{CFT} =
\delta S_{BH}$ and the solution is
\begin{eqnarray}
\delta E_L &=& \frac{2 \ell (2 M^2 - Q^2 - P^2) M}{Q (Q^2 + P^2)} \delta M - \frac{\ell (2 M^2 - Q^2 - P^2)}{Q^2 + P^2} \delta Q,
\nonumber\\
\delta E_R &=& \frac{2 \ell (2 M^2 - Q^2 - P^2) M}{Q (Q^2 + P^2)} \delta M - \frac{2 \ell M^2}{Q^2 + P^2} \delta Q.
\end{eqnarray}
These conjugate charges are actually identical with $\omega_L$ and $\omega_R$ by identifying $\delta M = \omega$ and $\delta Q = q$, namely
\begin{equation}
\tilde\omega_L = \delta E_L(\delta M = \omega, \delta Q = q), \qquad \tilde\omega_R = \delta E_R(\delta M = \omega, \delta Q = q).
\end{equation}
Moreover, one can also straightforwardly verify the following
relation for the imaginary part of the coefficient $c$,
\begin{equation}
2 \pi \gamma = \frac{\tilde\omega_L}{2 T_L} + \frac{\tilde\omega_R}{2 T_R}.
\end{equation}
Then the absorption cross section ultimately can be expressed as
\begin{equation}\label{CFTp}
P_\mathrm{abs} \sim T_L^{2h_L - 1} T_R^{2h_R - 1} \sinh\left( \frac{\tilde\omega_L}{2 T_L} + \frac{\tilde\omega_R}{2 T_R} \right) \left| \Gamma\left( h_L + i \frac{\tilde\omega_L}{2 \pi T_L} \right) \right|^2 \, \left| \Gamma\left( h_R + i \frac{\tilde\omega_R}{2 \pi T_R} \right) \right|^2,
\end{equation}
which has the same form of the finite temperature absorption cross
section of an operator with the conformal weights ($h_L, h_R$), frequencies ($\omega_L, \omega_R$),
electric charges ($q_L, q_R$) and chemical potentials ($\mu_L,
\mu_R$) in the dual 2D CFT with the temperatures ($T_L, T_R$).

%%%%%%%%%%%%%%%%%%%%%%%%%%%%%%%%%%%%%%%%%%%%%%%%%%%%%%%%%%%%%%%%%%%%%%
\section{Real time Correlator}
%%%%%%%%%%%%%%%%%%%%%%%%%%%%%%%%%%%%%%%%%%%%%%%%%%%%%%%%%%%%%%%%%%%%%%
Furthermore, since the real time correlator or the retarded Green's
function are of physical causal meaning, it is of importance to
calculate them in the AdS/CFT correspondence~\cite{Son:2002sd}. For
the Kerr black hole, this has been studied in~\cite{Chen:2010ni,
Becker:2010jj, Becker:2010dm}, so it is very interesting to
calculate the real time correlator from the perspective of the
RN/CFT$_2$ duality.

From the gravity side, note that the asymptotic behavior of the
ingoing scalar field~(\ref{asymR}) indicates that two coefficients
play different roles: $A$ as the source and $B$ as the response,
thus the two-point retarded correlator is simply~\cite{Chen:2010ni}
\begin{equation}
G_R \sim \frac{B}{A} = \frac{\Gamma(-2 l - 1)}{\Gamma(2 l + 1)} \, \frac{\Gamma(a) \, \Gamma(b)}{ \Gamma(c-a) \, \Gamma(c-b)}.
\end{equation}
Together with the identity~(\ref{cab}), we can easily check that
retarded Green function is
\begin{equation}
G_R \sim
%\frac{\Gamma(1 - 2 h)}{\Gamma(2 h - 1)}
\frac{\Gamma\left( h_L - i \frac{\tilde\omega_L}{2 \pi T_L} \right) \Gamma\left( h_R - i \frac{\tilde\omega_R}{2 \pi T_R} \right)}{\Gamma\left( 1 - h_L - i \frac{\tilde\omega_L}{2 \pi T_L} \right) \Gamma\left( 1 - h_R - i \frac{\tilde\omega_R}{2 \pi T_R} \right)}.
\end{equation}
Using the relation $\Gamma(z) \Gamma(1 - z) = \pi/\sin(\pi z)$ we have
\begin{eqnarray}\label{GreenR}
G_R &\sim& \sin\left(\pi h_L + i \frac{\tilde\omega_L}{2 T_L} \right) \sin\left(\pi h_R + i \frac{\tilde\omega_R}{2 T_R} \right)
\nonumber\\
&& \Gamma\left( h_L - i \frac{\tilde\omega_L}{2 \pi T_L} \right) \Gamma\left( h_L + i \frac{\tilde\omega_L}{2 \pi T_L} \right) \Gamma\left( h_R - i \frac{\tilde\omega_R}{2 \pi T_R} \right) \Gamma\left( h_R + i \frac{\tilde\omega_R}{2 \pi T_R} \right).
\end{eqnarray}
Moreover, since the conformal weights $h_L = h_R = l + 1$ are integers so
\begin{equation}
\sin\left(\pi h_L + i \frac{\tilde\omega_L}{2 T_L} \right) \sin\left(\pi h_R + i \frac{\tilde\omega_R}{2 T_R} \right) = (-)^{h_L + h_R} \sin\left(i \frac{\omega_L - q_L \mu_L}{2 T_L} \right) \sin\left(i \frac{\omega_R - q_R \mu_R}{2 T_R} \right).
\end{equation}

From the 2D CFT side, the Euclidean correlator, in terms of the
Euclidean frequencies $\omega_{EL} = i \omega_L$ and $\omega_{ER} = i
\omega_R$, is
\begin{eqnarray}
G_E(\omega_{EL}, \omega_{ER}) &\sim& T_L^{2h_L -1} T_R^{2h_R - 1} \mathrm{e}^{i \frac{\tilde\omega_{EL}}{2 T_L}} \mathrm{e}^{i \frac{\tilde\omega_{ER}}{2 T_R}}
\nonumber\\
&& \Gamma\left( h_L \!-\! \frac{\tilde\omega_{EL}}{2 \pi T_L} \right) \Gamma\left( h_L \!+\! \frac{\tilde\omega_{EL}}{2 \pi T_L} \right) \Gamma\left( h_R \!-\! \frac{\tilde\omega_{ER}}{2 \pi T_R} \right) \Gamma\left( h_R \!+\! \frac{\tilde\omega_{ER}}{2 \pi T_R} \right),
\end{eqnarray}
where
\begin{equation}
\tilde\omega_{EL} = \omega_{EL} - i q_L \mu_L, \qquad \tilde\omega_{ER} = \omega_{ER} - i q_R \mu_R.
\end{equation}
The retarded Green function $G_R(\omega_L, \omega_R)$ is analytic on
the upper half of the complex $\omega_{L,R}$-plane and it is related
to the Euclidean correlator by
\begin{equation}
G_E(\omega_{EL}, \omega_{ER}) = G_R(i \omega_L, i \omega_R), \qquad \omega_{EL}, \; \omega_{ER} > 0,
\end{equation}
and the Euclidean frequencies $\omega_{EL}$ and $\omega_{ER}$ should
take discrete values of the Matsubara frequencies at finite
temperature
\begin{equation}
\omega_{EL} = 2 \pi m_L T_L, \qquad \omega_{ER} = 2 \pi m_R T_R,
\end{equation}
in which $m_L, m_R$ are integers for bosons and half integers for
fermions. At these frequencies, the retarded Green function matches
well with the gravity side computation~(\ref{GreenR}) up to a
normalization factor depending on the charges $q_L$ and $q_R$, i.e.
the electric charge of the probe scalar field.

%%%%%%%%%%%%%%%%%%%%%%%%%%%%%%%%%%%%%%%%%%%%%%%%%%%%%%%%%%%%%%%%%%%%%%
\section{Conclusion}
%%%%%%%%%%%%%%%%%%%%%%%%%%%%%%%%%%%%%%%%%%%%%%%%%%%%%%%%%%%%%%%%%%%%%%
According to the well-known property that the near horizon (near)
extremal geometry of the 4D RN black hole only contains an AdS$_2$
structure, thus
%it is straightforward to reduce it to 2D asymptotically AdS$_2$ spacetime and
it is natural to believe that the associated holographic dual CFT
should be one-dimensional based on the idea of the usual
AdS$_D$/CFT$_{D-1}$ correspondence. The CFT$_2$ description is
expected to appear only until the 4D RN black hole is embedded into
five-dimensional spacetime where part of the $U(1)$ gauge potential
in 4D is transformed to the Kaluza-Klein vector as off-diagonal
components in the uplifted 5D metric, and then a (warped) AdS$_3$
geometry comes out in the near horizon (near) extremal limit.
Naively, by combining the uplifted and reduced perspectives, the 4D
RN black hole and its uplifted 5D counterpart seem to provide a
testable example for studying CFT$_1$ from CFT$_2$ by a simple
dimensional reduction.

However, in this paper, from the technique of probing the hidden
conformal symmetry of black hole backgrounds via external fields, we
suggest that even for the generic non-extremal 4D dyonic RN black
hole, it should still dual to a 2D CFT once we correctly incorporate
the contribution of the background gauge field. In practice, the
$U(1)$ symmetry of the background electromagnetic field can be
probed by an external charged scalar field, consequently, the hidden
2D conformal symmetry of the 4D RN black hole is revealed. In
mathematical language, the AdS$_3$ structure of the 4D RN black hole
nevertheless is encoded in the fiber bundle where the near horizon
(near) extremal AdS$_2$ geometry serves as the base manifold while
the $U(1)$ gauge field acts as the fiber. Hence a dimensional
reduction of the spacetime does not change the full symmetry
structure, but merely transforms part of the geometric information
from the base manifold into the fiber and vice versa. We investigate
all the technical details to read out the dual CFT$_2$ information
of the dyonic RN black hole from a probe charged scalar field and
show that the charged scalar field can reveal equivalent CFT$_2$
information probed by a neutral scalar field scattering in the 5D
uplifted RN black hole. Our results indicate a dyonic RN/CFT$_2$
correspondence and the approach can be further generalized to more
generic charged black holes. An interesting example is the
Kerr-Newman black hole we can explore, besides a recently
well-studied angular momentum AdS$_3$/CFT$_2$ description (which we
call it the $J$-picture), there should exist another dual CFT$_2$
description in which the $U(1)$ symmetry of the background
electromagnetic field can be probed (which we call it the
$Q$-picture). We would like to report the results in a forthcoming
paper~\cite{Chen:2010yw}.

\begin{appendix}
%%%%%%%%%%%%%%%%%%%%%%%%%%%%%%%%%%%%%%%%%%%%%%%%%%%%%%%%%%%%%%%%%%%%%%
\section{Symmetry and Casimir Operator of $\mathrm{AdS}_3$}
%%%%%%%%%%%%%%%%%%%%%%%%%%%%%%%%%%%%%%%%%%%%%%%%%%%%%%%%%%%%%%%%%%%%%%
In this Appendix, we summarized the basic properties of the AdS$_3$ space which are useful for our study in this paper. The metric of the AdS$_3$ space with radius $L$, in the Poincar\'e coordinates: ($w^\pm, y$), is
\begin{equation}\label{dsAdS3}
ds_3^2 = \frac{L^2}{y^2} ( dy^2 + dw^+ dw^-).
\end{equation}
There are two sets of symmetry generators
\begin{eqnarray}
H_1 = i \partial_+, \quad H_0 = i \left( w^+ \, \partial_+ + \frac12 y \, \partial_y \right), \quad H_{-1} = i \left( (w^+)^2 \, \partial_+ + w^+ y \, \partial_y - y^2 \, \partial_- \right),
\\
\bar H_1 = i \partial_-, \quad \bar H_0 = i \left( w^- \, \partial_- + \frac12 y \, \partial_y \right), \quad \bar H_{-1} = i \left( (w^-)^2 \, \partial_- + w^- y \, \partial_y - y^2 \, \partial_+ \right),
\end{eqnarray}
assembling two copies of the $SL(2,R)$ Lie algebra
\begin{equation}
\left[ H_0, H_{\pm 1} \right] = \mp i H_{\pm 1}, \qquad \left[ H_{-1}, H_1 \right] = - 2 i H_0.
\end{equation}
Thus the corresponding Casimir operator is
\begin{equation}
\mathcal{H}^2 = \bar\mathcal{H}^2 = - H_0^2 + \frac12 \left( H_1 H_{-1} + H_{-1} H_1 \right) = \frac14 \left( y^2 \, \partial_y^2 - y \, \partial_y \right) + y^2 \, \partial_+ \partial_-.
\end{equation}

The Poincar\'e metric (\ref{dsAdS3}) can be transformed to a
black hole metric with coordinates ($t, r, \chi$) via the following
coordinate transformations
\begin{eqnarray}\label{wy2tr}
w^+ &=& \sqrt{\frac{r - r_+}{r - r_-}} \, \exp(2 \pi T_R \chi + 2 n_R t),
\nonumber\\
w^- &=& \sqrt{\frac{r - r_+}{r - r_-}} \, \exp(2 \pi T_L \chi + 2 n_L t),
\nonumber\\
y &=& \sqrt{\frac{r_+ - r_-}{r - r_-}} \, \exp[\pi (T_R + T_L) \chi + (n_R + n_L) t],
\end{eqnarray}
we can directly calculate all the $SL(2,R)$ generators in terms of black hole coordinates
\begin{eqnarray*}
H_1 &=& i \mathrm{e}^{-(2 \pi T_R \chi + 2 n_R t)} \left[ \sqrt{\Delta} \partial_r \!+\! \frac{n_L (\delta_- \!+\! \delta_+) \!+\! n_R (\delta_- \!-\! \delta_+)}{4 \pi \sqrt{\Delta} \mathcal{A}} \partial_\chi \!-\! \frac{T_L (\delta_- \!+\! \delta_+) \!+\! T_R (\delta_- \!-\! \delta_+)}{4 \sqrt{\Delta} \mathcal{A}} \, \partial_t \right],
\nonumber\\
H_0 &=& i \left[ \frac{n_L}{2 \pi \mathcal{A}} \, \partial_\chi - \frac{T_L}{2 \mathcal{A}} \, \partial_t \right],
\nonumber\\
H_{-1} &=& i \mathrm{e}^{2 \pi T_R \chi + 2 n_R t} \left[ -\sqrt{\Delta} \partial_r \!+\! \frac{n_L (\delta_- \!+\! \delta_+) \!+\! n_R (\delta_- \!-\! \delta_+)}{4 \pi \sqrt{\Delta} \mathcal{A}} \partial_\chi \!-\! \frac{T_L (\delta_- \!+\! \delta_+) \!+\! T_R (\delta_- \!-\! \delta_+)}{4 \sqrt{\Delta} \mathcal{A}} \, \partial_t \right],
\end{eqnarray*}
and
\begin{eqnarray*}
\bar H_1 &=& i \mathrm{e}^{-(2 \pi T_L \chi + 2 n_L t)} \left[ \sqrt{\Delta} \partial_r \!-\! \frac{n_R (\delta_- \!+\! \delta_+) \!+\! n_L (\delta_- \!-\! \delta_+)}{4 \pi \sqrt{\Delta} \mathcal{A}} \partial_\chi \!+\! \frac{T_R (\delta_- \!+\! \delta_+) \!+\! T_L (\delta_- \!-\! \delta_+)}{4 \sqrt{\Delta} \mathcal{A}} \, \partial_t \right],
\nonumber\\
\bar H_0 &=& i \left[ - \frac{n_R}{2 \pi \mathcal{A}} \, \partial_\chi - \frac{T_R}{2 \mathcal{A}} \, \partial_t \right],
\nonumber\\
\bar H_{-1} &=& i \mathrm{e}^{2 \pi T_L \chi + 2 n_L t} \left[ - \sqrt{\Delta} \partial_r \!-\! \frac{n_R (\delta_- \!+\! \delta_+) \!+\! n_L (\delta_- \!-\! \delta_+)}{4 \pi \sqrt{\Delta} \mathcal{A}} \partial_\chi \!+\! \frac{T_R (\delta_- \!+\! \delta_+) \!+\! T_L (\delta_- \!-\! \delta_+)}{4 \sqrt{\Delta} \mathcal{A}} \, \partial_t \right],
\end{eqnarray*}
where
\begin{equation}\label{deltaA}
\delta_\pm = r - r_\pm, \qquad \mathcal{A} = T_R n_L - T_L n_R.
\end{equation}
Finally the Casimir operator becomes
\begin{equation}\label{Casimir}
\mathcal{H}^2 = \partial_r \Delta \partial_r - \frac{r_+ - r_-}{r - r_+} \left( \frac{T_L + T_R}{4 \mathcal{A}} \partial_t - \frac{n_L + n_R}{4 \pi \mathcal{A}} \partial_\chi \right)^2 + \frac{r_+ - r_-}{r - r_-} \left( \frac{T_L - T_R}{4 \mathcal{A}} \partial_t - \frac{n_L - n_R}{4 \pi \mathcal{A}} \partial_\chi \right)^2.
\end{equation}

\end{appendix}

%%%%%%%%%%%%%%%%%%%%%%%%%%%%%%%%%%%%%%%%%%%%%%%%%%%%%%%%%%%%%%%%%%%%%%
\section*{Acknowledgement}
%%%%%%%%%%%%%%%%%%%%%%%%%%%%%%%%%%%%%%%%%%%%%%%%%%%%%%%%%%%%%%%%%%%%%%
This work was supported by the National Science Council of the R.O.C. under the grant NSC 96-2112-M-008-006-MY3 and in part by the National Center of Theoretical Sciences (NCTS).

%%%%%%%%%%%%%%%%%%%%%%%%%%%%%%%%%%%%%%%%%%%%%%%%%%%%%%%%%%%%%%%%%%%%%%

\end{document}